\documentclass[fleqn,usenatbib]{mnras}

\usepackage{newtxtext,newtxmath}

\usepackage[T1]{fontenc}

\DeclareRobustCommand{\VAN}[3]{#2}
\let\VANthebibliography\thebibliography
\def\thebibliography{\DeclareRobustCommand{\VAN}[3]{##3}\VANthebibliography}

\usepackage{graphicx}	
\usepackage{amsmath}	
\usepackage[normalem]{ulem}
\usepackage{dirtytalk}
\usepackage{hyperref}
\usepackage[dvipsnames]{xcolor}
\usepackage[nameinlink, capitalise, noabbrev]{cleveref}

\defcitealias{Wen2015}{WH15}
\defcitealias{Sabater2019}{S19}

\title[Environmental star formation suppression]{Local versus global environment: the suppression of star formation in the vicinity of galaxy clusters}

\author[K. de Vos et al.]{
K. de Vos,$^{1}$\thanks{E-mail: ppykd1@nottingham.ac.uk (KdV)}
M. R. Merrifield,$^{1}$
N. A. Hatch,$^{1}$
\\
$^{1}$School of Physics and Astronomy, University Park, University of Nottingham, Nottingham, NG7 2RD, UK}

\date{Accepted XXX. Received YYY; in original form ZZZ}

\pubyear{2015}

\begin{document}
\label{firstpage}
\pagerange{\pageref{firstpage}--\pageref{lastpage}}
\maketitle

\begin{abstract}
In order to examine where, how and why the quenching of star formation begins in the outskirts of galaxy clusters, we investigate the de-projected radial distribution of a large sample of quenched and star-forming galaxies (SFGs) out to $30R_{500}$ around clusters. We identify the SFG sample using radio continuum emission from the Low-Frequency Array Two-metre Sky Survey. We find that the SFG fraction starts to decrease from the field fraction as far out as $10R_{500}$, well outside the virial radius of the clusters. We investigate how the SFG fraction depends on both large-scale and local environments, using radial distance from a cluster to characterise the former, and distance from 5th nearest neighbour for the latter. The fraction of SFGs in high-density local environments is consistently lower than that found in low-density local environments, indicating that galaxies' immediate surroundings have a significant impact on star formation. However, for high-mass galaxies -- and low mass galaxies to a lesser extent -- high-density local environments appear to act as a protective barrier for those SFGs that survived this pre-processing, shielding them from the external quenching mechanisms of the cluster outskirts. For those galaxies that are not in a dense local environment, the global environment causes the fraction of SFGs to decrease toward the cluster centre in a manner that is independent of galaxy mass.  Thus, the fraction of SFGs depends on quite a complex interplay between the galaxies' mass, their local environment, and their more global cluster-centric distance.  

\end{abstract}

\begin{keywords}
surveys –- galaxies: clusters: general -– galaxies: evolution -– galaxies: groups: general -– galaxies: star formation
\end{keywords}



\section{Introduction}
\label{sec:Introduction}

There has long been evidence to suggest a strong correlation between the star formation rate (SFR) of a galaxy, and the density of the environment in which it lives \citep{Oemler1974, Davis1976, Dressler1980, Dressler1983, Postman1984, Balogh1997, Poggianti1999, Peng2010}. Star-forming galaxies (SFGs) are seen more commonly in lower-density field environments, and quiescent galaxies are more commonly seen in denser environments such as groups or clusters. This is often referred to as the SFR-density relation, and emphasises the importance of environmental effects on the processes that drive galaxy evolution.

This relation strongly suggests that a galaxy's transition from a low-density environment into a high-density one involves external physical processes that reduce or even extinguish entirely the star-formation within a galaxy, which is a process known as quenching. However, there are many suggested external mechanisms that contribute to quenching, and it is likely that various combinations of them are responsible for shutting down the star formation in a galaxy, depending on the environments that a galaxy encounters throughout its life. These various phenomena include: \say{strangulation} or \say{starvation}, which is the process by which a galaxy is no longer able to accrete cold gas onto its disk, thereby preventing any further star formation from occurring \citep{Larson1980}; ram pressure stripping, whereby a galaxy's cold, star-forming gas is pushed out due to the relative velocity difference between its host galaxy and the hot, dense, virialised gas of the medium surrounding it \citep{Gunn1972, Abadi1999}; \say{harassment}, which are tidal interactions disruptive to star-formation, caused by the gravitational pull of other nearby galaxies in high density environments \citep{Farouki1981, Moore1996}; and galaxy mergers, which have a similar effect \citep{Makino1997, Angulo2009, Wetzel2009a, Wetzel2009b, White2010, Cohn2012}.

Additionally, a galaxy's star-formation may be affected before even arriving at a cluster, during the period within which it is travelling towards a cluster along filaments or in groups. This process is know as pre-processing, and is the means by which a galaxy experiences quenching prior to crossing the virial radius, due to quenching mechanisms associated with higher density local environments than the field. Pre-processing is thought to be a large contributor to the high proportion of quiescent galaxies in clusters \citep{Zabludoff1998, McGee2009}, and could be an explanation for some of the observed timescales that suggest quenching starts prior to infall \citep{Wetzel2013, Haines2015, Werner2021}.

The size of a cluster is generally referred to in terms of its virial radius, which is the radius within which all mass is mixed and relaxed. The virial radius is comparable to $R_{200}$, which is the radius within which the average density is two hundred times the critical density of the universe, $\rho_c$. $R_{500}$ is an analogously-defined radius which follows the same logic in its definition, where $R_{500}\sim 0.7R_{200}$ \citep{Navarro1996}. There is a great deal of literature that investigates quenching processes for infalling satellite galaxies out to $\sim 2-5R_{200}$ \citep{Balogh1998, Lewis2002, Bahe2013, Wetzel2014, Haines2015, Bianconi2018, Pintos-Castro2019, Lacerna2022, Baxter2022, Salerno2022, Hough2023, Kesebonye2023, Rihtarsic2023, Lopes2024}, and any radii further than this is commonly considered the field. However, \citet{Haines2015} found that the fraction of galaxies that are star-forming still had not reached the observed field fraction by $3R_{200}$, suggesting that the influence of the cluster outskirts on the star-forming properties of galaxies must extend further. Indeed, \citet{deVos2021} found that the gaseous medium in cluster outskirts had a strong influence on the morphology of  radio galaxies as far out as $10R_{500}$. More recently, \citet{Lopes2024} have shown that quenching begins as far out as $5R_{200}$ from the cluster centre, and that group galaxies have an undeniably lower fraction of SFGs than isolated galaxies, which provides strong evidence for pre-processing. 

To extend this analysis, we corroborate and expand on these exciting new results: we investigate this transitionary region in the far cluster outskirts in order to further quantify the distance from the cluster centre that the SFG fraction starts to deviate from the SFG field fraction value. In addition, we investigate the effects of both intrinsic and external galaxy properties on the SFG fraction by measuring the influence of both stellar mass and distance to nearest neighbour, in order to determine the various mechanisms that could be influencing quenching at the distances seen. In \cref{sec:Data_Method}, we describe the data and method used, while \cref{sec:results} presents the resulting fractional distributions with respect to cluster-centric radius, and discusses the various possibilities for the physical mechanisms responsible for them. Finally, in \cref{sec:summary}, we summarise the somewhat surprising results.

\section{Data and Method}
\label{sec:Data_Method}

\subsection{Data and sample selection}
\label{subsec:data_sample_selec}

To conduct this study, we use data from the Low-Frequency Array (LOFAR) Two-metre Sky Survey (LoTSS) second data release \citep[DR2,][]{Shimwell2022}, the Sloan Digital Sky Survey (SDSS) sixteenth data release \citep[DR16,][]{Ahumada2020}, the MPA-JHU value-added catalogue \citep{Brinchmann2004b}, and the Wide-field Infrared Survey Explorer (WISE) allWISE source catalogue \citep{Wright2010}.

LoTSS DR2 \citep{Shimwell2022} is a 6" resolution, low-frequency radio survey spanning 27\% of the Northern Sky over two large regions. The catalogue is comprised of over 4 million radio sources observed at a central frequency of 144MHz, with a median rms sensitivity of 83 $\mu$ Jy beam$^{-1}$. LoTSS is an excellent survey with which to identify star-forming galaxies, as its 120-168Mhz observations are ideal for detecting the low-frequency radio synchrotron emission given off by supernovae cosmic rays. This tracer of star-formation has a high level of completeness, as it is not victim to dust extinction in the same way that H$\alpha$ emission is. Furthermore, radio-calculated star-formation rates (SFRs) are more accurate in their ability to account for the entire on-sky extended source, as opposed to SFR calculations done with single-fiber H$\alpha$ measurements from the centre of the galaxy.

The optical counterparts to the radio data for this study are selected from the sources in the SDSS DR16 \citep{Ahumada2020} spectroscopic catalogue defined as being in the class \say{galaxy}\footnote{\url{https://www.sdss4.org/dr16/spectro/spectro_access/}}. These optical sources are complemented by emission line fluxes, stellar masses and star formation rates derived from the emission line analysis of SDSS DR7 by MPA-JHU \citep{Brinchmann2004b}, as well as infrared magnitude values collected in the allWISE survey \citep{Wright2010}.

Two master catalogues are produced from the above source tables. The first comprises the spectroscopic galaxy catalogue from SDSS DR16, with additional data from MPA-JHU value-added catalogue and the allWISE source catalogue matched within 2" on the sky. This results in a catalogue of over 3 million optically-identified galaxies with spectroscopic redshifts, for which 85\% have WISE magnitude data, and 27\% have emission line fluxes from MPA-JHU. As only 23\% of sources in this catalogue have good stellar mass values from MPA-JHU, we calculate a stellar mass proxy, $M_{i}$, for all galaxies in this sample based on their i-band magnitudes, $i$, using the best fit formula 
\begin{equation}
\log_{10}(M_{i}) = -0.42i + 1.93.
\end{equation}

This sample will hereafter be referred to as \textit{SDSS galaxies}. The second master catalogue, \textit{radio galaxies}, comprises the radio sources from LoTSS DR2 matched within 2" on the sky to \textit{SDSS galaxies}. For SFGs, such a simple matching process reliably pairs up the optical and radio emission.

Finally, we utilise the cluster catalogue from \citet{Wen2012} and \citet[hereafter \citetalias{Wen2015}]{Wen2015} in order to determine the global environments of the galaxies in this sample. The \citetalias{Wen2015} cluster catalogue defines a sample of 158,103 clusters within the redshift range $0.02<z\leq 0.8$, identified using a friends-of-friends algorithm from galaxies in the SDSS DR12 catalogue \citep{Alam2015}. The catalogue has a false detection rate of $<6\%$, and is 95\% complete for clusters of mass $M_{200} > 10^{14} M_\odot$ (where $M_n$ is the mass inside radius $R_n$, within which the density is $n$ times the critical density of the Universe). \citetalias{Wen2015} define the position of the brightest cluster galaxy (BCG) to be the cluster centre, and identifies the BCG as the brightest galaxy within $\pm 0.04(1 + z)$ and 0.5\,Mpc of the densest region of each cluster.

\textit{SDSS galaxies}, \textit{radio galaxies} and the \citetalias{Wen2015} cluster catalogue are limited to the same region of sky and the redshift range $0.05<z\leq 0.2$. We implement a lower redshift limit of $z > 0.05$ as nearby galaxies are resolved into multiple emission regions due to the telescope's 6 arcsec resolution. This means that, below this lower limit, we are unable to match a single LOFAR source with a single optical galaxy. The upper redshift limit at $z < 0.2$ is implemented to minimise issues arising from completeness and galaxy evolution. To ensure that both samples are directly comparable in their mass limits at all redshifts, we divide the \textit{radio galaxies} sample into redshift bins of width 0.01, and apply a mass threshold that cuts off at 95\% of each bin's sample size. We then apply the same threshold to \textit{SDSS galaxies}. This adopted lower mass threshold corresponds to a mass range of (0.6--6.3)$\times 10^{10} M_\odot$ across the redshift range of our sample. Finally, only entries in the \citetalias{Wen2015} cluster catalogue with spectroscopic redshifts are kept, in order to avoid photometric uncertainty when matching galaxies and clusters. This results in a sample size of 213,072 sources for \textit{SDSS galaxies}, and 68,168 sources for \textit{radio galaxies}. 

\subsection{Classification of SFGs}
\label{subsec:SFG_classification}

In order to extract the sub-sample of SFGs from all sources of emission in \textit{radio galaxies}, we follow \citet[hereafter \citetalias{Sabater2019}]{Sabater2019} and \citet{Herpich2016} by using a simple WISE colour diagram cut at $W2 - W3 = 0.8$. \citetalias{Sabater2019} showed that this colour cut is effective at separating SFGs and active galactic nuclei (AGN). We therefore adopt this singular diagnostic method to classify the SFGs in this sample, which results in a sample of 50,516 radio-identified SFGs that will hereafter be referred to as \textit{radio SFGs}.

We calculate the SFRs, $\psi$ of \textit{radio SFGs} following the SFR-$L_{150}$ relation outlined in \citet{Smith2021},
\begin{equation}
    1.058\log_{10}\psi = \log_{10}L_{\rm{150MHz}} - 22.221,
\end{equation}
where $L_{\rm{150MHz}}$ is the radio luminosity at 150MHz. Specific star formation rates (sSFR) for \textit{radio SFGs} is then given by $\psi/M_{i}$. The minimum sSFR threshold is set at $4\times10^{-12}$ yr$^{-1}$.

\subsection{Cluster matching}
\label{subsec:cluster_matching}

To quantify each galaxy's relationship to a cluster environment, we associate objects in \textit{SDSS galaxies} and \textit{radio SFGs} with the closest cluster in the \citetalias{Wen2015} cluster catalogue by following a similar method to that used in \citet{Garon2019} and adopted in \citet{deVos2021}. For each individual cluster at redshift $z_{clus}$, we identify the most likely galaxies associated with that cluster using the formula
\begin{equation}
    \frac{|z_{gal}-z_{clus}|}{1+z_{gal}}<0.04,
\end{equation}
where $z_{gal}$ is the galaxy redshift. Of these remaining galaxies, only those within a radially projected distance on the sky of $<50r_{500}$ are matched to the cluster. At this preliminary stage, some objects are matched to multiple clusters. There is no way to determine for individual objects with which cluster they are most closely associated, but the process described in \cref{subsec:bg_subtraction} allows us to resolve this issue statistically.

\subsection{Nearest neighbour matching}
\label{subsec:nn_matching}

In order to determine a measure of the local environment of each galaxy, we conduct a nearest neighbour matching process. To minimise line-of-sight contamination, we restrict the search for neighbours to galaxies that were assigned to the same cluster in \cref{subsec:cluster_matching}. This is done by conducting an on-sky match of \textit{SDSS galaxies} to itself for each galaxy's 3rd, 5th and 10th nearest neighbour. The offset, $\theta$, between each galaxy and its nearest neighbours is then used to calculate the physical on-sky distance between them, using $d_{nn} = \theta D$, where $D$ is the angular diameter distance found using each cluster's redshift. These same nearest neighbour distances are then applied to the matched galaxies in \textit{radio SFGs}. One concern might be that some of these distances are distorted by edge effects in the survey, but we find that the impact is negligible out to beyond $40R_{500}$, which lies at the largest radii considered in this study (see \cref{subsec:bg_subtraction} below).

In the subsequent analysis, we use $d_{5nn}$, but the results are qualitatively similar for $d_{3nn}$ and $d_{10nn}$.

\subsection{Background subtraction}
\label{subsec:bg_subtraction}

\begin{figure}
    \centering
    \includegraphics[width=\columnwidth]{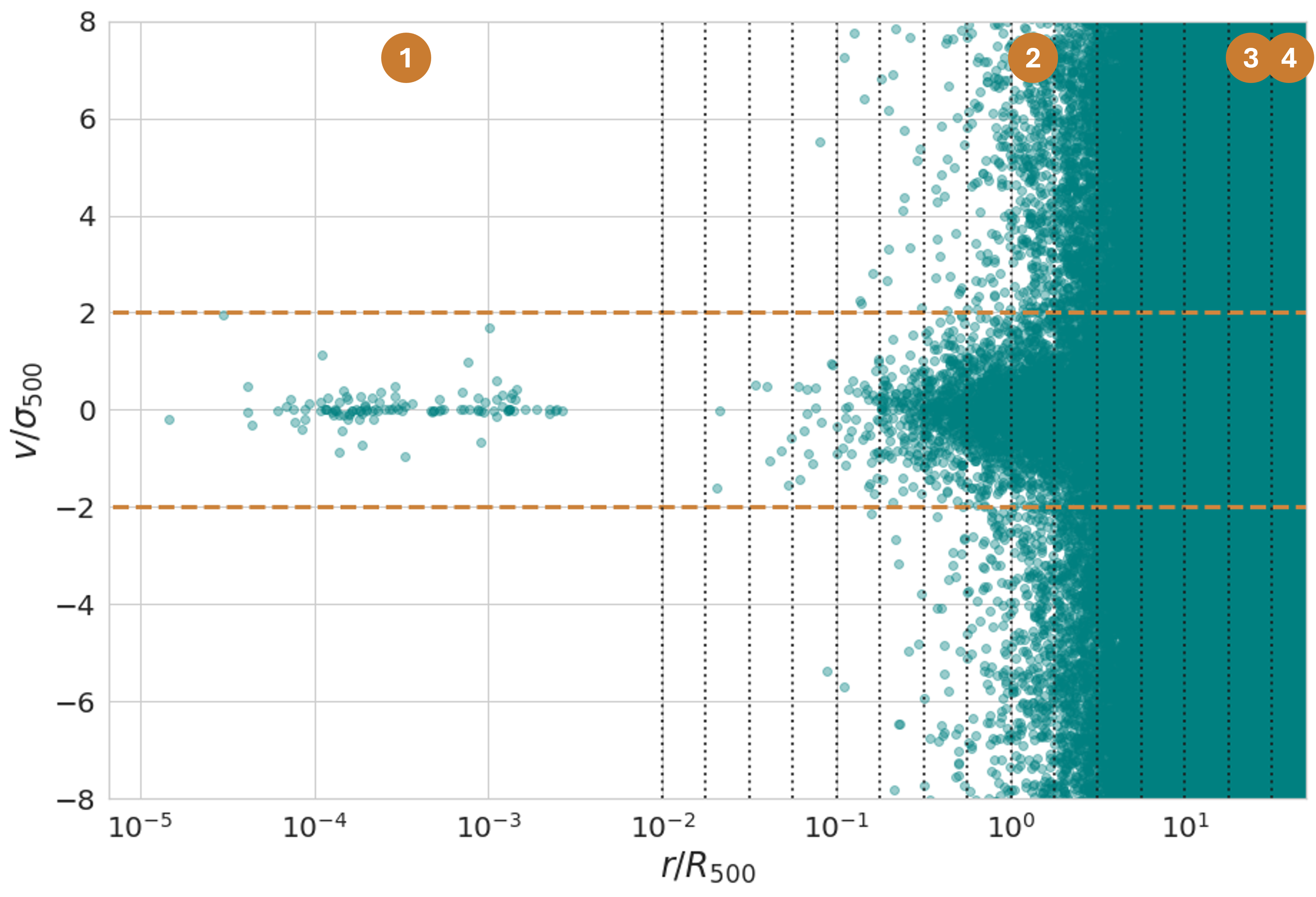}
    \caption{The projected phase-space distribution (showing line-of-sight velocity versus projected separation) for \textit{radio SFGs} out to $50r_{500}$. The orange, horizontal, dashed lines at $v = \pm 2\sigma_{500}$ indicate the limits of velocity associated with the cluster, and the black, vertical, dotted lines indicate the cluster radius slicing used to determine the line-of-sight contamination in each annulus bin. The numbered, orange circles correspond to the histogram bins shown in \cref{fig:Background_hists}.}
    \label{fig:Phase_space}
\end{figure}

Since the density of galaxies physically associated with a cluster decreases with radius, the membership assigned in \cref{subsec:cluster_matching} will see increasing contamination with radius.  It is not possible to determine which individual galaxies are truly associated with each cluster, but we can correct for this contamination statistically. In order to assess this contamination, we analyse the phase space for both \textit{SDSS galaxies} and \textit{radio SFGs} individually, the latter of which is shown in \cref{fig:Phase_space}. In order to combine data from multiple clusters, each galaxy's distance from their host cluster and associated velocity is scaled by their cluster's $r_{500}$ and characteristic velocity $\sigma_{500}$ respectively.

\begin{figure}
    \centering
    \includegraphics[width=\columnwidth]{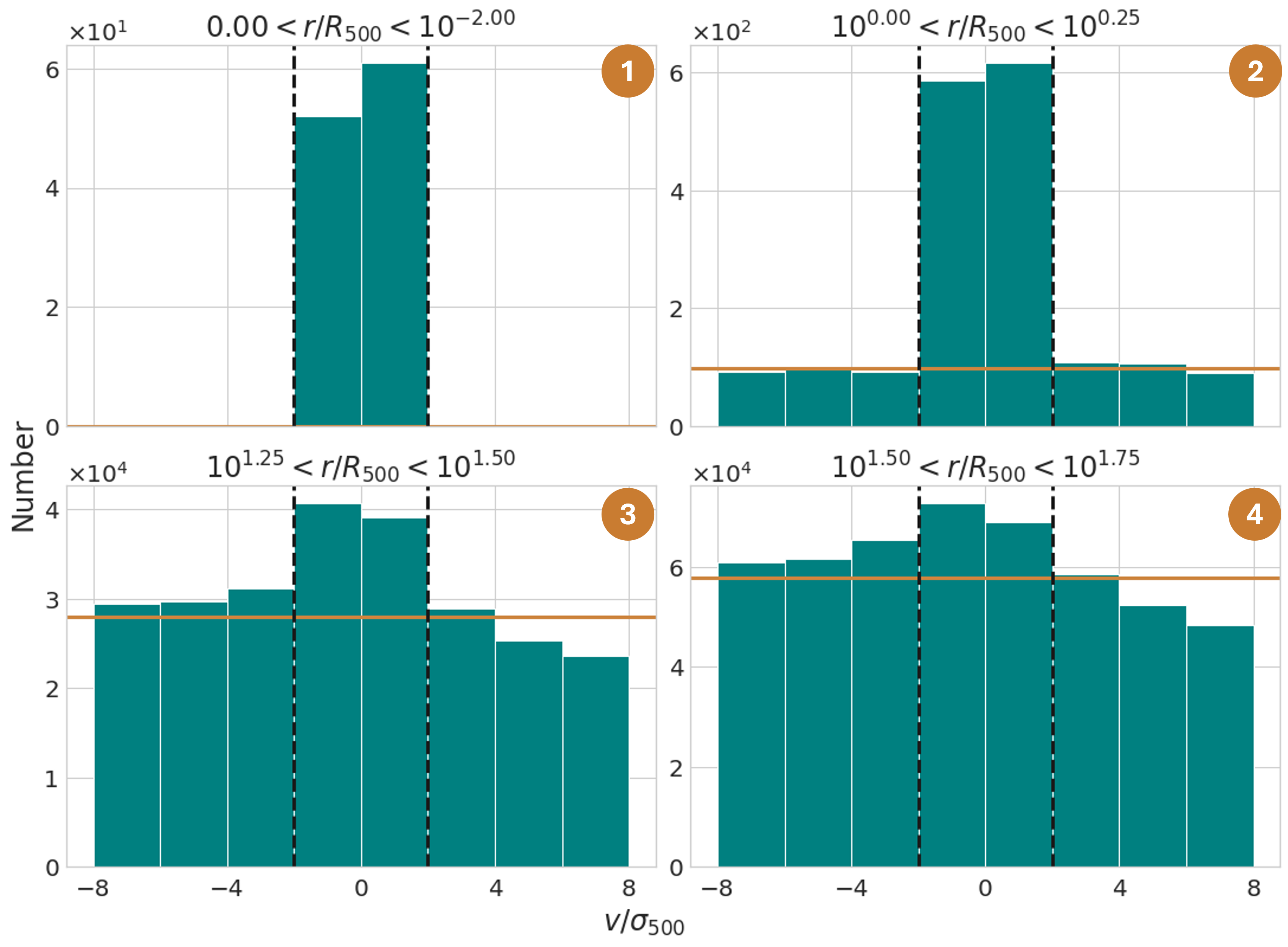}
    \caption{The line-of-sight velocity of galaxies within four of the logarithmically-spaced annuli from \cref{fig:Phase_space}. Galaxies within the black, vertical, dashed lines (indicating the characteristic velocity range $v = \pm 2\sigma_{500}$) are deemed to be correctly associated with their host cluster, and those outside are identified as contaminants. The orange, horizontal, solid line shows the mean of the background contamination in each bin, and the numbered, orange circles correspond to the annulus bins shown in \cref{fig:Phase_space}.}
    \label{fig:Background_hists}
\end{figure}

It is clear from \cref{fig:Phase_space} that there is a signal of correctly-associated galaxies lying within the characteristic velocity range of $v = \pm 2\sigma_{500}$, but this signal becomes significantly contaminated outside of $\sim R_{500}$. By slicing this phase space up into logarithmically-spaced radial bins, we are able to ascertain the level of line-of-sight contamination in each individual annulus, as illustrated in \cref{fig:Background_hists}.

In each bin, the line-of-sight contamination is determined using the mean of the number of galaxies outside of $v = \pm 2\sigma_{500}$, as indicated by the horizontal line in \cref{fig:Background_hists}. There is a slight difference in background level on the two sides of the cluster due to the variation in completeness with redshift, so we take the average to define the background contamination at the cluster redshift. This value is subtracted from the signal of cluster-associated galaxies within $v = \pm 2\sigma_{500}$ to give the background-subtracted signal in each annulus. These values are then divided by the annulus area to give the number density, $N(R)$, for each radial bin. Finally, we take the error on the number density in each annulus to be the Poisson error, and adopt the field value to be the number density of background galaxies within all annuli.

It is interesting to note that this cluster-associated signal seen in \cref{fig:Background_hists} remains statistically-significant out to the penultimate radial bin at more than $30 R_{500}$, which for a typical cluster in this sample is more than $20\,{\rm Mpc}$. This indicates that the overdensity of galaxies associated with a cluster extends to many times the virial radius. 

\subsection{Abel inversion}

The quantities that we derive in \cref{subsec:bg_subtraction} measure projected densities on the plane of the sky, $N(R)$, but these quantities are intrinsically over-predicted towards the centre of the cluster due to projection. To determine more intrinsic physical quantities, we need to de-project these values into volume densities. Since the process of stacking many clusters together averages out any asymmetries in structure, we can assume spherical symmetry in the density profiles, so the projected densities can be straightforwardly converted to spatial density via an Abel inversion, using
\begin{equation}
    \label{eq:Abel_Inv}
    n(r) = -\frac{1}{\pi}\int_{r}^{\infty}\frac{dN}{dR}\frac{1}{(R^2-r^2)^{\frac{1}{2}}}dR,
\end{equation}
where $n(r)$ and $r$ are the de-projected number density and radial distance respectively, and $N(R)$ and $R$ are the projected equivalent variables. In the numerical implementation of this quadrature, the value in the outermost bin considered is indeterminate. Fortunately, here we can use the overall average field value from the entire survey as a boundary condition at this point. The imposition of this constraint has the desirable effect of suppressing the error amplification inherent in such a de-projection.  Finally, we can calculate errors on the de-projected distributions by applying a Monte Carlo method that resamples the projected data using their Poisson errors. 

\begin{figure}
    \centering
    \includegraphics[width=\columnwidth]{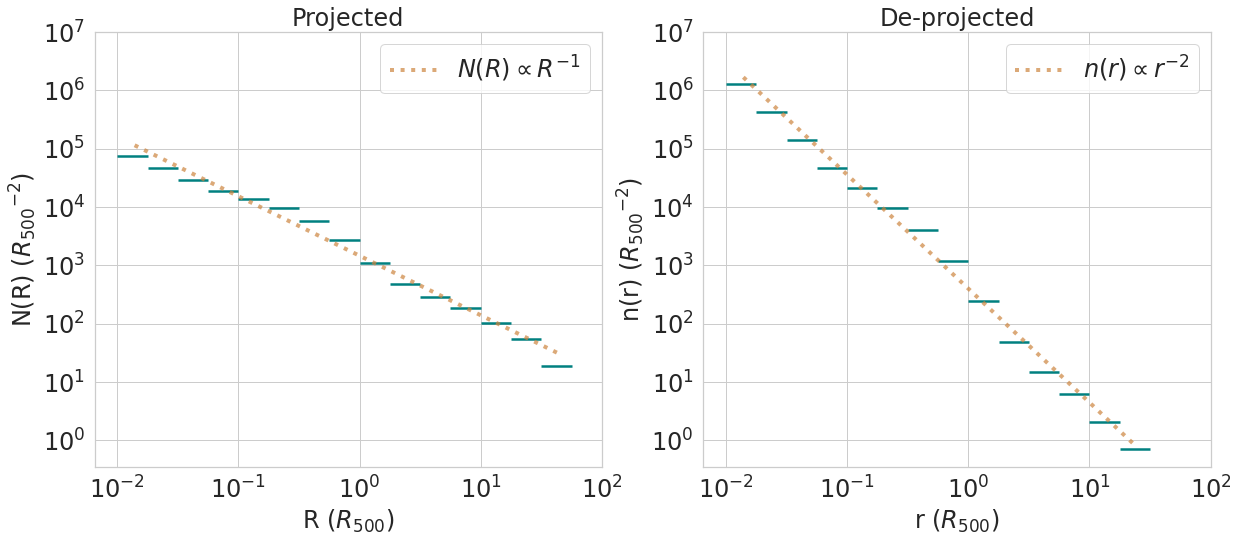}
    \caption{The projected and de-projected (via an Abel inversion) number density distributions of \textit{SDSS galaxies}, with respect to cluster radius. The dotted lines show the power laws that fit to these distributions.}
    \label{fig:Number_density}
\end{figure}

By way of illustration, \cref{fig:Number_density} shows the projected number density profile of the full \textit{SDSS galaxies} sample with projected distance from their associated cluster, and the numerical implementation of the Abel inversion, which  yields the variation in number density with radius. As previously noted by \citet{Beers1986}, the projected number density of galaxies around clusters follows an approximate power law of $N(R) \propto R^{-1}$. The numerically-derived inversion yields a number density in good agreement with the analytic solution to Equation \cref{eq:Abel_Inv} that exists in this case, given by $n(r) \propto r^{-2}$. This overall power law agrees well with the general profile determined by \citet{Beers1986}, although more recent analyses have indicated that the interplay between luminous and dark matter may lead to some level of departure from such a simple power law \citep{Newman2013}.

\section{Results}
\label{sec:results}

\subsection{Total SFG fraction}
\label{subsec:total_SFG_frac}

\begin{figure}
    \centering
    \includegraphics[width=\columnwidth]{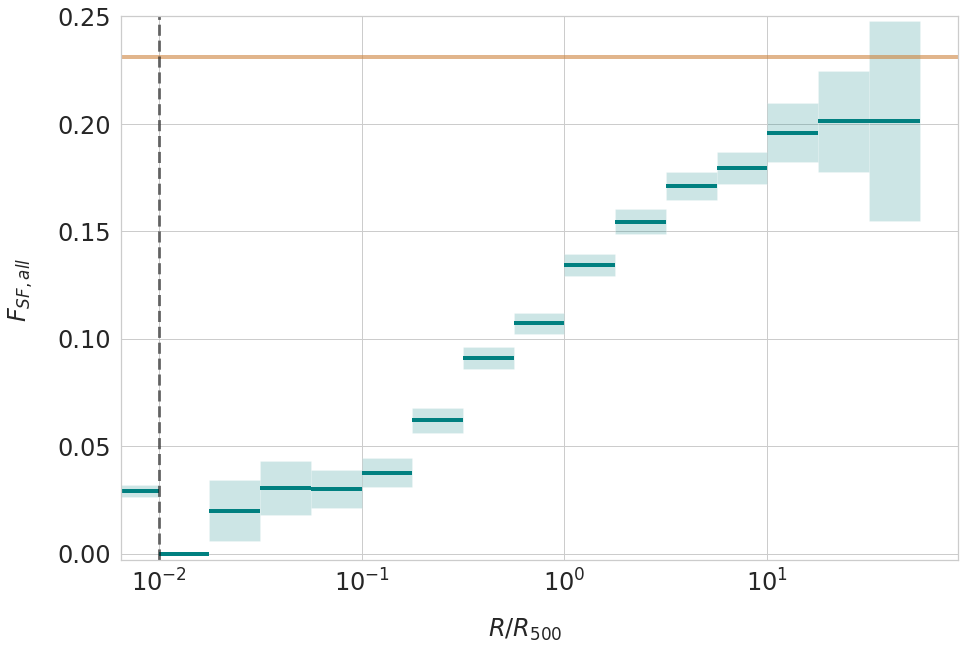}
    \caption{The projected distribution of the fraction of galaxies identified as star-forming, with respect to cluster radius. The orange, horizontal, solid line is a measure of the field fraction of SFGs (see \cref{subsec:SFG_classification}), and the black, vertical, dashed line represents the radius at which everything within is considered to be the very centre of the galaxy cluster. This inner bin includes all galaxies between $0 < r < 0.01R_{500}$.}
    \label{fig:Gal_frac_proj}
\end{figure}

Using the two catalogues described in \cref{subsec:data_sample_selec}, \textit{SDSS galaxies} and \textit{radio SFGs}, we can determine how the fraction of galaxies that are star-forming changes with respect to distance from the centre of the associated cluster. This projected fraction is shown in \cref{fig:Gal_frac_proj}, where $F_{SF,all}=N_{SF}/N_{SDSS}$, and $N_{SF}$ and $N_{SDSS}$ are the number densities of \textit{radio SFGs} and \textit{SDSS galaxies} respectively. The SFG field fraction is also shown, but it is worth noting that \textit{radio SFGs} only comprises 21\% of \textit{SDSS galaxies}, as opposed to the $\sim60\%$ \citep{Dressler1980} seen when selecting SFG samples via H-$\alpha$ emission. This difference reflects the higher threshold that these radio data place on the detection of star formation. Our measured field SFG fraction is similar to that found in other studies of high mass galaxies, but lower than expected for the lower mass galaxies in our sample \citep{Corcho-Caballero2020}. Nonetheless, since all of the comparisons in this work are made in a relative sense, the absolute value of this fraction has no impact on any of the results obtained.

\begin{figure}
    \centering
    \includegraphics[width=\columnwidth]{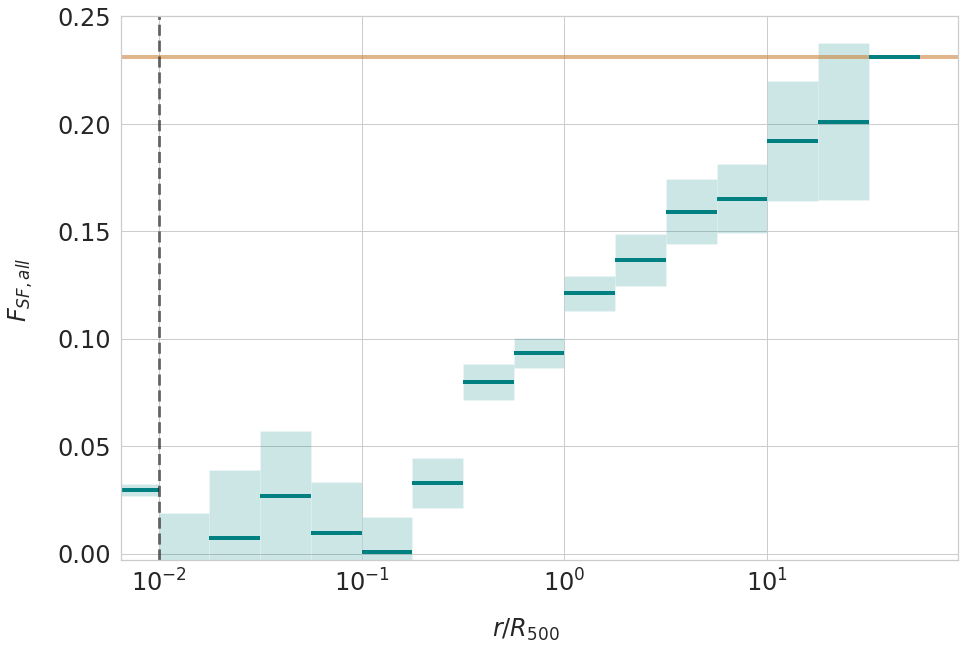}
    \caption{The de-projected distribution of \cref{fig:Gal_frac_proj}.}
    \label{fig:Gal_frac_AI}
\end{figure}

It is immediately apparent from this figure that there is a steady decline in the fraction of SFGs with decreasing radius, starting at at least $10R_{500}$, if not further. Since projection has the effect of combining data from a range of radii, this plot will tend to average away some of the true variation with radius. Indeed, the Abel-inverted spatial density presented in \cref{fig:Gal_frac_AI} shows an even more dramatic decline in the star-forming fraction, which reaches zero within the central $0.1R_{500}$.  For the remainder of this paper we therefore generally only present the de-projected results as representative of the intrinsic properties of the systems, and due to the small sample size of any SFGs in annuli $<10^{-0.75}R_{500}$, we make the decision to only investigate how the SFG fraction changes for galaxies at radii of $>10^{-0.75}R_{500}$ for the remaining analyses.
 
At the large radii of $\sim 10 R_{500}$ where the fraction of SFGs starts to drop, we are unlikely to be seeing cluster members or even the "backsplash" galaxies that have recently fallen through the cluster and have yet to virialize, as these objects seldom travel more than $\sim 3R_{500}$ from the cluster centre \citep{Haggar2020}. Thus the decrease in star-forming galaxies does seem simply to reflect the properties of galaxies on their first infall onto the cluster. To investigate what the driving mechanism for this transition might be, we next explore whether it is driven by an internal property such as mass or an external one like the local density around the galaxy.

\subsection{Stellar mass}
\label{subsec:stell_mass}

\begin{figure}
    \centering
    \includegraphics[width=\columnwidth]{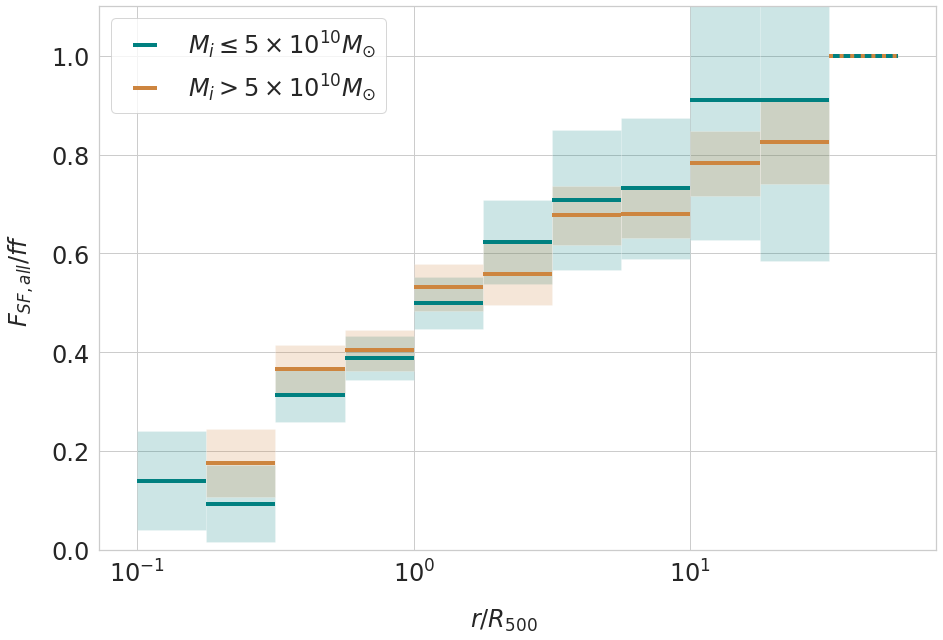}
    \caption{The de-projected distribution of the fraction of SFGs that fall into the two mass bins either side of $M_{i} = 5\times 10^{10}M_{\odot}$, normalised by their respective field fractions: 0.24 for the low mass bin, and 0.19 for the high mass bin.}
    \label{fig:GF_stellmass_AI}
\end{figure}

In order to quantify any dependence on the mass of the galaxy, we split both samples - \textit{radio SFGs} and \textit{SDSS galaxies} - into two roughly equally-sized bins at the stellar mass value $M_{i} = 5\times 10^{10}M_{\odot}$, and determine how the fraction of SFGs in each sub-sample varies with cluster-centric radius. There is a lower fraction of high-mass SFGs than low-mass SFGs, therefore in order to easily compare the difference in the two distributions, each fraction is normalised by its field fraction value, which is 0.24 for $M_{i} \leq 5\times 10^{10}M_{\odot}$ and 0.19 for $M_{i} > 5\times 10^{10}M_{\odot}$. The resulting radial profiles are presented in \cref{fig:GF_stellmass_AI}.

Immediately, we see that the decline still begins at very large radii for both mass populations, suggesting that the distribution seen in \cref{fig:Gal_frac_AI} is not being dominated by one mass population over another. On closer inspection, we notice that there are slight differences in the two sample distributions; an example being that the gradient for the decline of the low-mass sample appears slightly steeper and starts declining later than the high-mass sample. However, as these slight differences are within errors and therefore not statistically significant, it is hard to draw any concrete conclusions from these minor differences. As such, we conclude that the mass of the galaxy does not seem to be a significant factor in determining how or why star-formation is being suppressed at such large radii. We therefore now turn to local environment, to see if this can be held responsible.

\subsection{Local environment}
\label{subsec:local_env}

In order to investigate how the local environments of SFGs affect their fractional distribution with respect to cluster radius, we separate \textit{radio SFGs} into six bins based on each galaxy's distance to its 5th nearest neighbour, $d_{5nn}$. This parameter $d_{5nn}$ is defined such that higher values correspond to lower local densities, and vice-versa.

\begin{figure*}
    \centering
    \includegraphics[width=\textwidth]{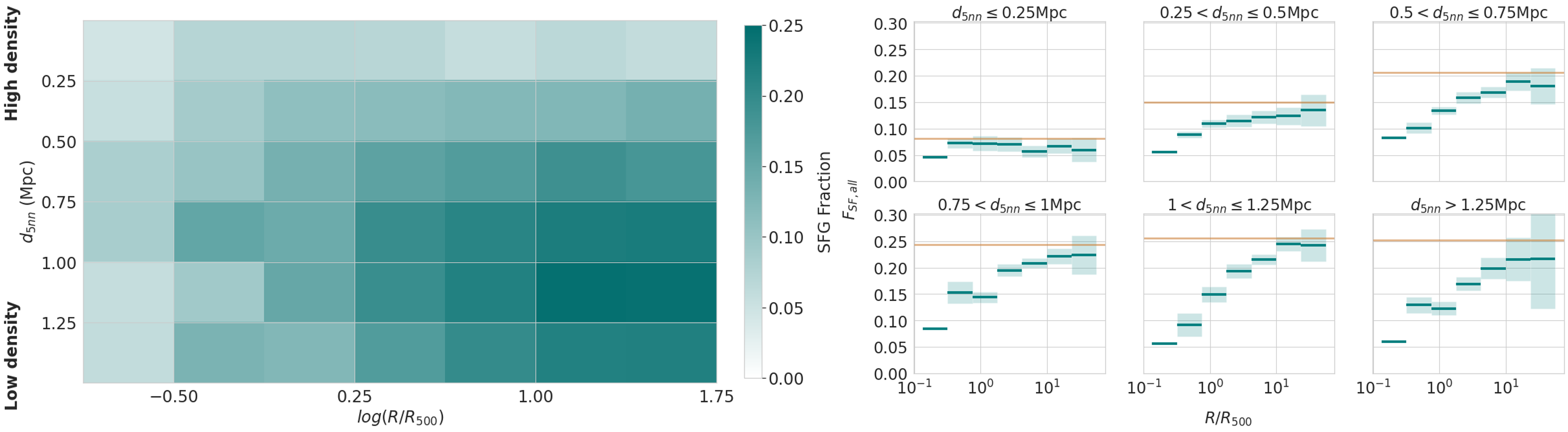}
    \caption{\textit{Left:} The projected tile distribution of the fraction of galaxies identified as star-forming, with respect to the distance to the 5th nearest neighbour galaxy and the $R_{500}$-normalised distance from the cluster centre. \textit{Right:} The associated projected distributions of the fraction of galaxies identified as star-forming with respect to cluster radius, binned by the distance to the 5th nearest neighbour galaxy. The horizontal, orange lines are a measure of the field fraction for each bin. Each subplot equates to a row in the left-hand tile plot.}
    \label{fig:nn5_comb}
\end{figure*}

\begin{figure*}
    \centering
    \includegraphics[width=\textwidth]{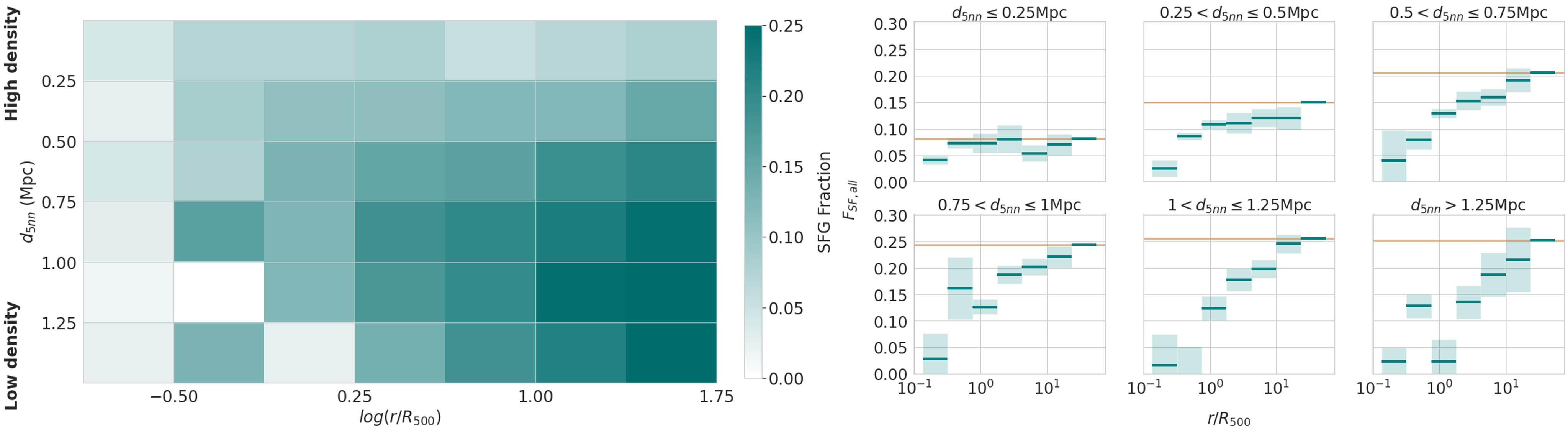}
    \caption{The de-projected distributions of \cref{fig:nn5_comb}.}
    \label{fig:nn5_comb_AI}
\end{figure*}

When dividing the data into so many bins, the noise-amplifying effects of the Abel inversion become significant, so in this case we present both the observed projected distributions (\cref{fig:nn5_comb}) and the de-projected physical quantities (\cref{fig:nn5_comb_AI}). We also present the results as both profiles for individual density bins in which the size of the uncertainty is explicitly presented (right panels) and as a two-dimensional tile plot of radius and local density, which gives a clearer overall impression of how the star-forming fraction depends on both variable (left panels).  Comparing these two figures, it is clear that the main features seen are not an artefact of the deprojection process.

When analysing the right-hand plots of these figures, we see a largely flat distribution in the high density local environment bins that gradually steepens into a declining distribution as density decreases. The higher local density distributions also have a lower field SFG fraction, which increases with decreasing density. 

We interpret the combination of these results as follows: SFGs that become part of higher density regions such as groups and filaments in the galaxy cluster outskirts undergo pre-processing which quenches some, but not all SFGs \citep{Zabludoff1998, McGee2009}. This suppression of star-formation results in a lower overall SFG fraction in these types of local environments, but those environments then appear to host a flatter distribution with radius than that seen in the lower density local environments too. This smaller variation with radius implies that higher density local environments shield their surviving SFGs from any global environment mechanisms that are responsible for the declining trend we see in lower density bins, perhaps because the inter-galaxy medium is dense enough to deflect gas associated with the cluster from further stripping cold, star-forming gas from these objects, thus resulting in the flattened distribution. This potential shielding would be less and less effective at lower densities which explains the gradual steepening seen as local environment density decreases.

These trends are also apparent in the left panels of \cref{fig:nn5_comb,fig:nn5_comb_AI}, with the extra information that we can more readily compare the absolute value of the SFG fraction at different local densities.  It is apparent from these plots that the chances of a SFG surviving during its infall into a cluster is a balance of two factors: at high local densities, pre-processing by this local environment quenches most SFGs; at low local densities, the influence of the cluster starts to win out at larger radii, also quenching SFGs; so it is at intermediate local densities that the fraction of SFGs stays high to the smallest radii. However, in all cases eventually the cluster wins, quenching almost all SFGs at small radii, irrespective of their local environment \citep{Choque-Challapa2019,Haggar2023}.

Having found this interplay between local and global environment, the remaining question is whether the protection afforded by the local environment depends on the intrinsic property of the mass of the galaxy in a subtle way that wasn't apparent when we just considered galaxy mass in \cref{subsec:stell_mass}.

\subsection{Stellar mass vs local environment}
\label{subsec:sm_vs_le}

\begin{figure}
    \centering
    \includegraphics[width=\columnwidth]{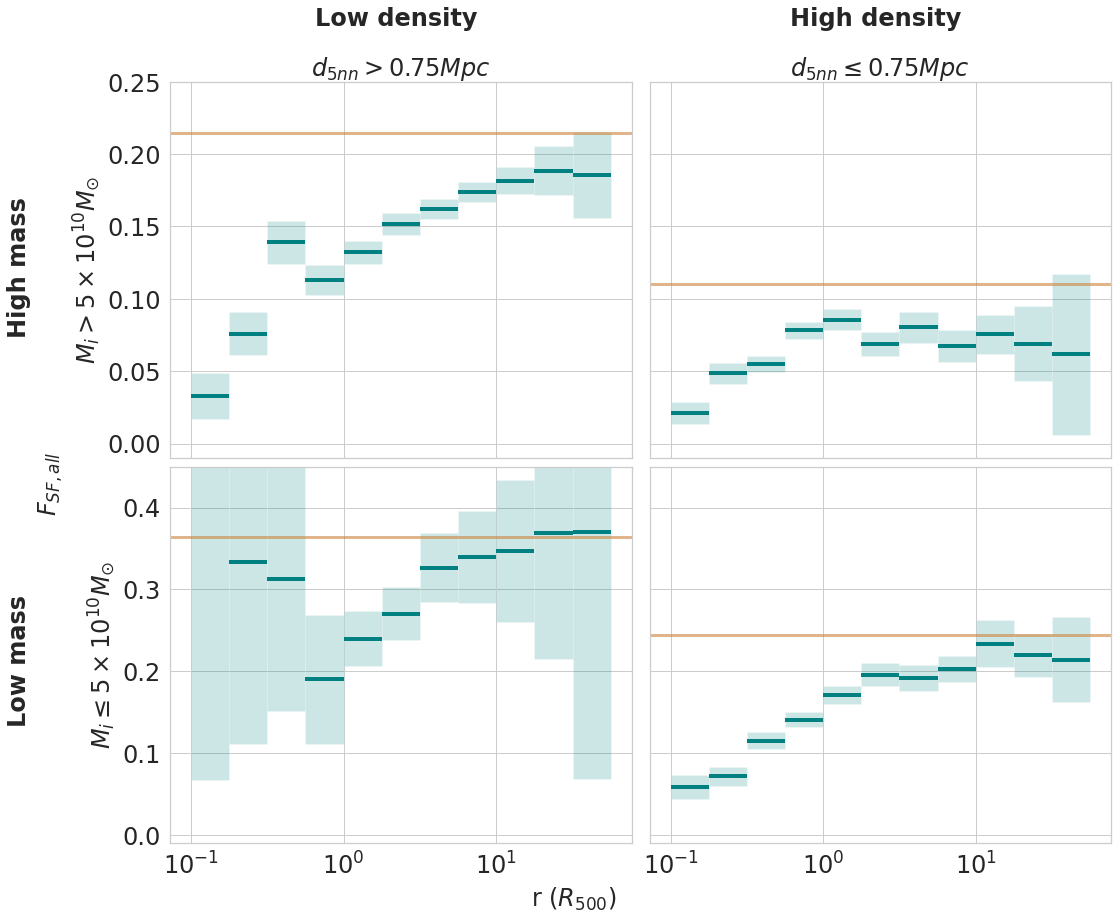}
    \caption{The projected distribution of the SFG fraction, binned by both $M_{i}$ and $d_{5nn}$. The binning boundaries are given by $M_{i} = 5\times 10^{10}M_{\odot}$ and $d_{5nn} = 0.75$Mpc, and the horizontal, orange lines are a measure of the field fraction for each bin.}
    \label{fig:mass_nn5}
\end{figure}

Finally, we analyse how the variation in radial distribution that we see when binning by local environment density, might also be mass-dependent. We do this by separating both of the \textit{radio SFGs} and \textit{SDSS galaxies} samples into four bins, split by low and high stellar mass either side of $M_{i} = 5\times 10^{10}M_{\odot}$, and low and high local environment density either side of $d_{5nn} = 0.75$Mpc. The results of this analysis are shown in \cref{fig:mass_nn5}. As in \cref{subsec:local_env}, we show the projected distribution here, but very similar results are seen in the de-projected distributions.

We find declining cluster-centric SFG fractions similar to the trends seen in \cref{fig:Gal_frac_proj,fig:Gal_frac_AI} in both of the low-density subsamples. The high-density, high-mass sample, however, exhibits a flat distribution at large radii, down to $\sim 1r_{500}$. The final subsample of high-density, low-mass galaxies shows a similar trend to that seen for its high-mass counterpart, but is somewhat less clear-cut and could therefore also be interpreted as a shallow decline. This distinction suggests that the flattening of the cluster-centric SFG fraction exhibited in the high local environment density plots of \cref{fig:nn5_comb,fig:nn5_comb_AI} is predominantly driven by the subsample of higher mass galaxies, and to a lesser extent by lower mass galaxies.

One possible interpretation of the results in \cref{fig:mass_nn5} may be that the lower fraction of massive (and possibly also low-mass) SFGs that are able to continue forming stars -- despite the pre-processing effects of their higher local density -- are then protected from further loss of star formation due to the approaching cluster, by that same denser local environment. This protection appears to hold until $1R_{500}$, at which point the SF fraction decline begins again -- presumably, past this point the defence provided by a higher-density local environment is not enough to shield its remaining SFGs from quenching. This subtle effect was hidden in \cref{fig:GF_stellmass_AI} when the samples were only divided by stellar mass and local environment was not considered. In this more nuanced analysis, however, we see that the stellar mass of a galaxy \textit{does} impact the effectiveness of the global environment on galaxy quenching, but only when the galaxy resides in a particular dense local environment.

\section{Summary \& Discussion}
\label{sec:summary}

Through this analysis, we show that there is a consistent and persistent trend displaying a decline in the SFG fraction with decreasing cluster-centric radius, that begins by at least $10R_{500}$ from the cluster centre - even further than the $5R_{200}$ ($\sim7R_{500}$) found in the recent, similar results of \citet{Lopes2024}. We interpret this decline as the quenching of the SFG population prior to infall into a cluster, as the large radius at which this decline begins is too far out to be explained by backsplash galaxies, which only travel out to $\sim3R_{500}$ at most \citep{Haggar2020}. In order to determine what might be driving this decline in the SFG fraction, we investigate how the radial distribution is affected when binning by stellar mass, distance to nearest neighbour (local environment density), and a combination of both.

When binning by stellar mass, we find that both populations still experience this same decline at large radii, past $10R_{500}$, and experience very little difference in the gradient of their decline. It is discussed that there is - unsurprisingly - a lower fraction of high mass SFGs than low mass ones, but the rate at which they are both reduced is comparable. If this decline is due to quenching as predicted, then these results imply that all SFGs, regardless of mass, experience the same rate of quenching outside of $R_{500}$. Consequently, whatever quenching mechanism(s) are responsible for this decline are independent of mass, and galaxy mass is clearly not solely responsible for the SFG fraction decline that we see at such large radii.

We then turn to binning by distance to the 5th nearest neighbour, $d_{5nn}$, in order to determine if local environment is the driver behind the declining trend we see at large radii. As it happens, we do in fact observe that local environment affects the rate at which the star-forming fraction declines, with denser local environments experiencing less SFG quenching, and less-dense local environments experiencing more. We also find that the overall fraction of SFGs is lower in denser environments, which suggests that they likely have gone through some pre-processing upon joining a higher density environment. Additionally, we observe that upon approach to the cluster centre, there is a lower fraction of SFGs across the board, irrespective of the density of their local environment. These findings therefore suggest that the declining trend seen in the total SFG population is primarily due to the SFGs in lower density local environments experiencing quenching at large radii. We interpret the lack of a radial trend for the SFGs in higher density environments to be due to their local environment providing a form of protection from quenching by the global environment. These results are in agreement with those seen in \citet{Lopes2024}.

Finally, we conclude this analysis by attempting to determine whether the seemingly protective qualities seen from high local environment densities depend on the stellar mass of the galaxies, in a more nuanced way that was not visible in \cref{subsec:stell_mass}. We find that this is in fact the case in higher density regions, and that the distribution flattening seen in galaxies residing in high density local environments (see \cref{subsec:local_env}) primarily applies to high mass galaxies, with a less pronounced but still detectable shielding effect influencing lower-mass galaxies.

We postulate that these results suggest one of two possibilities in high density local environments. The first being that these high density regions, such as groups and filaments, are mainly only capable of shielding high mass galaxies from any potential global environment mechanisms responsible for quenching in the cluster outskirts. Low mass galaxies may be more susceptible to these mechanisms, and therefore the shielding effect of the high density environment is not always strong enough to protect them. Alternatively, we could be seeing that high density local environments \textit{are} effective at shielding all SFGs, irrespective of mass, from global environment quenching mechanisms, but they are in fact host to quenching mechanisms of their own that are more effective on low-mass galaxies, thus leaving the high mass SFG population unaffected. 

This second possibility could provide an explanation for the driver behind the declining distribution seen here in high density local environments, however it still does not explain the driver behind the universally declining distributions seen in low density regions for galaxies of all masses. Nevertheless, these findings suggest that whatever mechanism(s) are responsible for the decline of the SFG fraction must be mass-independent in low density local environments, and potentially \textit{mass-dependent} in high density local environments. Ultimately, it would be very interesting to see the specific mechanisms responsible for these results identified and associated with their respective quenching effects, in both high and low local environment densities, in order to understand the root of the quenching of SFGs on the outskirts of galaxy clusters.

\section*{Acknowledgements}

KdV, NAH and MRM acknowledge support from the UK Science and Technology Facilities Council (STFC) under grant ST/X000982/1.
All authors are grateful for the use of data from LOFAR, the LOw Frequency ARray, and SDSS, the Sloan Digital Sky Survey.
This research made use of Astropy, a community-developed core Python package for astronomy \citep{astropy:2013, astropy:2018} hosted at \url{http://www.astropy.org/}, of MATPLOTLIB \citep{Hunter:2007}, of Plotly \citep{plotly}, and of TOPCAT \citep{Taylor:2005}.

\section*{Data Availability}

All of the individual catalogues used in this paper are pubicly available. The LoTSS DR2 data can be found at \url{https://www.lofar-surveys.org/releases.html}, the SDSS DR16 data can be found at \url{https://www.sdss4.org/dr16/spectro/spectro_access/}, the MPA-JHU data can be found at \url{https://wwwmpa.mpa-garching.mpg.de/SDSS/DR7/}, and the allWISE data can be found at \url{https://irsa.ipac.caltech.edu/cgi-bin/Gator/nph-scan?mission=irsa&submit=Select&projshort=WISE}. The cluster catalogues by \citet{Wen2012} and \citet{Wen2015} are also publicly available and are associated with the referenced papers. For access to the cluster-matched data compiled from these catalogues, please contact KdV.



\bibliographystyle{mnras}
\bibliography{bibliography} 


\bsp	
\label{lastpage}
\end{document}